\documentclass[12pt]{article}
\usepackage{jheppub}

\usepackage{setspace}

\usepackage[usenames,dvipsnames,table]{xcolor}
\usepackage{graphicx,amsmath,amssymb,multirow,array,bm,mathrsfs}
\usepackage{epsf,amsfonts}
\usepackage[numbers,sort&compress]{natbib}
\usepackage{pdfpages}

\usepackage{slashed}

\newcommand{\be}{\begin{equation}}
\newcommand{\ee}{\end{equation}}
\newcommand{\bea}{\begin{eqnarray}}
\newcommand{\eea}{\end{eqnarray}}

\newcommand{\OO}{{\cal O}}

\newcommand{\p}{\partial}

\newcommand{\DL}{{\Delta_L}}

\newcommand{\dxm}{{\Delta x^-}}
\newcommand{\dxp}{{\Delta x^+}}

\def\d{\delta}

\def\r{\rho}

\def\a{{\tilde \alpha}}

\def\d{\delta}

\def\G{\Gamma}

\def\G{\Gamma}

\title{Notes on AdS-Schwarzschild eikonal phase }
\author{Andrei Parnachev, Kallol Sen}
\affiliation{School of Mathematics and Hamilton Mathematics Institute,\\ Trinity College Dublin, Dublin 2, Ireland}
\emailAdd{parnachev, kallol [at] maths.tcd.ie}

\abstract{
We consider  the eikonal phase associated with the gravitational scattering of a highly energetic light particle off a very heavy
object in AdS spacetime.
A simple expression for this phase follows from the WKB approximation to the scattering amplitude and
has been computed to all orders in the ratio of the impact parameter to the Schwarzschild radius of the heavy particle.
The eikonal phase is related to the deflection angle  by the usual stationary phase relation.
We consider the flat space limit and observe that
for sufficiently small impact parameters (or angular momenta) the eikonal phase develops a large
imaginary part; the inelastic cross-section is exactly the classical absorption cross-section of the black hole.
We also consider a double scaling limit where the momentum becomes null simultaneously with the
asymptotically AdS black hole becoming very large.
In the dual CFT this limit retains contributions from all leading twist multi stress tensor operators, which are universal
with respect  to the addition of higher derivative  terms to the gravitational lagrangian.
We compute the eikonal phase and the associated Lyapunov exponent in the double scaling limit.
}

\begin{document}
\maketitle


\bigskip

\section{Introduction and Summary}
\subsection{Introduction}
Gravitational high-energy scattering probes an interesting regime of quantum gravity.
At large impact parameters (very small momentum transfer) the scattering amplitude 
is given by a single graviton exchange.
As the impact parameter is lowered, the amplitude is related by a Fourier transform in the
impact parameter space to the exponential of the eikonal phase (see e.g. \cite{Giddings:2011xs} for a  review).
As the impact parameter $b$ becomes comparable with the Schwarzschild radius $R_s$ associated with the total energy,
a black hole may form -- this is reflected in the imaginary part of the eikonal phase.
These issues have been a subject of active investigation starting from 
\cite{tHooft:1987vrq,Amati:1987wq,Muzinich:1987in,Sundborg:1988tb,Amati:1987uf,Amati:1990xe,Verlinde:1991iu}.
For example, ref. \cite{Giddings:2007qq} advocated a black hole ansatz to describe the breakdown of unitarity.
Generally one needs the full eikonal phase, to all orders in the ratio $R_s/b$, to study the black hole
regime.

A simplification happens when one of the particles is very heavy,
so that its mass is the largest scale in the problem, and the other particle is highly relativistic.
In this case the eikonal phase determines the deflection of null geodesics in the Schwarzschild background 
(see e.g.
 \cite{Kabat:1992tb,DAppollonio:2010krb,Neill:2013wsa,Akhoury:2013yua,Bjerrum-Bohr:2014zsa,Bjerrum-Bohr:2016hpa,Luna:2016idw,Cachazo:2017jef,Bjerrum-Bohr:2018xdl,Cheung:2018wkq,Kosower:2018adc,Bern:2019nnu,KoemansCollado:2019ggb,Bern:2019crd,Bjerrum-Bohr:2019kec,Damour:2019lcq,Bern:2020gjj,Blumlein:2020znm,Cheung:2020gyp,Cristofoli:2020uzm,Bini:2020wpo,Bern:2020buy,Parra-Martinez:2020dzs,
Kalin:2020mvi,DiVecchia:2020ymx,Damour:2020tta,Cheung:2020gbf} for  some work related 
 to the computation of the heavy-light scattering angle.)
The deflection angle can in principle be computed to all orders in $R_s/b$.

A similar problem can be posed in  AdS spacetime\footnote{
See e.g.   \cite{Cornalba:2006xk,Cornalba:2006xm,Cornalba:2007zb,Brower:2007qh,Cornalba:2007fs,Costa:2012cb,Camanho:2014apa,Kulaxizi:2017ixa,Li:2017lmh,Costa:2017twz,Meltzer:2019pyl}  for  work addressing the  eikonal phase in AdS.  }
, where
the eikonal phase (a.k.a. the phase shift) is an interesting object from the
point of view of the holographically dual CFT.
For example, in  \cite{Giusto:2020mup} the phase shift was used to distinguish black hole microstates from
conical defects.
In \cite{Kulaxizi:2018dxo} it was argued that conformal two-point functions  in a
generic heavy state  could be used to obtain the AdS-Schwarzschild eikonal phase.
Moreover, the phase  is only sensitive to the stress-tensor sector of the correlator,
which only contains contributions from the stress tensor and its composites.

There has been some progress in computing the stress-tensor sector of such correlators
 \cite{Fitzpatrick:2019zqz,Karlsson:2019qfi,Li:2019tpf,Kulaxizi:2019tkd,Fitzpatrick:2019efk,Karlsson:2019dbd,Li:2019zba,Karlsson:2019txu,Karlsson:2020ghx,Li:2020dqm,Parnachev:2020fna,Fitzpatrick:2020yjb}.
In particular, in \cite{Fitzpatrick:2019zqz} the leading twist stress tensor OPE coefficients in holographic CFTs
were shown to be largely universal -- independent of the higher derivative terms in the bulk gravitational action.
 In \cite{Kulaxizi:2019tkd} the contributions of all leading twist double stress tensors in such CFTs 
 were  shown to produce a very simple function.
 In \cite{Karlsson:2019dbd} it was explained how the leading twist stress tensor
 sector can be computed by bootstrap and in \cite{Karlsson:2020ghx} it was shown how to go beyond the leading twist -- 
 the phase shift played an important role in this story.
So, the AdS-Schwarzschild eikonal phase is an important object.
In this paper we discuss its properties and investigate various limits.

\subsection{Summary and outline}
As we review in Section 2, to compute the phase shift one needs to Fourier transform a heavy-heavy-light-light (HHLL)
correlator on the boundary.
The Fourier transformed correlator is a function of the energy, $p^t$, and the angular momentum $p^\varphi$.
In the  limit where $p^t$ and $p^\varphi$ are large, the eikonal phase can be computed exactly as a function of
the ratio $\alpha=p^\varphi/p^t$, related to the impact parameter, and is given by  \cite{Kulaxizi:2018dxo} 
\be
\label{psoper}
   \delta = p^t \Delta t - p^\varphi \Delta \varphi  \, ,
\ee
where $\Delta t$ and $\Delta \varphi$ are the time and angular displacements of the null geodesic
with the energy $p^t$ and angular momentum $p^\varphi$.

As we describe in Section 2, this is simply a consequence of the WKB approximation to the differential
equation which determines the holographic correlator.
Another way to illustrate how eq. (\ref{psoper}) emerges involves considering a 
probe particle with the mass which is large in AdS units.
The two-point function is  then determined by the length of  spacelike geodesics and is peaked around points
connected by null geodesics.
Such points form a codimension one subspace on the boundary.
There is a  null geodesic which gives the dominant contribution.
The parameters of this geodesic can be determined from the stationary phase condition -- it is precisely the
null geodesic whose energy and angular momentum are equal to the  $p^t$ and $p^\varphi$ 
parameters of the Fourier transform.
Hence, we end up with the expression (\ref{psoper}) for the phase shift.

The AdS eikonal phase is directly related to the usual eikonal phase in the flat space scattering computed in the probe limit.
To get the latter, one simply needs to take the flat space limit of the AdS result, which we do in Section 3.
One can confirm that the resulting formula is the conventional flat space eikonal phase 
 by considering the relation between the AdS phase shift and the deflection angle --
this relation is exactly the same as the one which follows from the stationary phase approximation to the eikonal
scattering amplitude in flat spacetimes.

The flat space limit yields a very simple formula for the eikonal phase, as we observe in Section 3.
It can be summed to give an analytic function; we perform this summation in four spacetime dimensions.
The resulting phase is real for impact parameters larger than the radius of the circular null orbit, but
develops a large imaginary part for smaller impact parameters.
As a result, the total inelastic cross-section is equal to the geometric absorption cross-section of the Schwarschild black hole.

In Section 4 we consider the opposite limit of  large impact parameters. 
More precisely, we take a double scaling limit where the impact parameter becomes large 
(the momentum approaches the lightcone)
and at the same time the Schwarzschild radius also becomes large in  AdS units.
This limit is particularly interesting from the dual CFT point of view -- it retains contribution from
all leading twist multi-stress tensors.
We study the propagation of null geodesics in the the effective metric derived in \cite{Parnachev:2020fna}
and use eq. (\ref{psoper}) to compute the phase shift. 
It agrees (as it should) with the corresponding limit of the full phase shift.
We also compute the Lyapunov exponent for the null geodesics approaching the null orbit
in the effective metric.

We discuss our results in Section 5. Appendices contain some technical details used in the main text.


\section{Eikonal phase  in AdS-Schwarzschild }

Ref. \cite{Kulaxizi:2018dxo} argued that  two-point functions
in certain heavy states in holographic CFTs with a large central charge $C_T$
can be used to define the eikonal phase (a.k.a. the phase shift).
One should simply Fourier transform the CFT correlator on the $d$-dimensional Lorentzian cylinder
(the boundary of the $d+1$ dimensional asymptotically AdS spacetime)
\be
\label{defps}
  e^{i\delta} \simeq  \int dt d\varphi    \ e^{-i p^t t } \ (\sin \varphi)^{d-2} 
        C_{p^\varphi}^{d-2\over2} (\cos \varphi) \langle  \OO_H \OO_L(t, \varphi)  \OO_L(0)  \OO_H \rangle \, ,
\ee 
where the heavy operators $\OO_H$ (with the conformal dimension $\Delta_H \sim C_T$) are inserted at $t=\pm\infty$ and $t,\varphi$ are
the displacements of the two light operators [with $\Delta_L \simeq  \OO(1) $]  on the cylinder ($\varphi$ is the relative angle on the $d-1$-dimensional spatial sphere of radius $R$ ).
In  (\ref{defps}) $C_{p^\varphi}^{d-2\over2} (\cos \varphi)$ are the Gegenbauer polynomials with the angular momentum $p^\varphi$ ,
 which generalize the spherical  harmonics of the $d=4$ case.
 The momenta are taken to be large, $p^t \gg R^{-1}, p^\varphi \gg1$ and the integral in (\ref{defps})
 can be computed in the stationary phase approximation.
 Substituting the large $p^\varphi$ behavior of the  Gegenbauer polynomials, (\ref{defps}) can be written as 
\be
\label{defpsa}
  e^{i\delta} \simeq  \int d t d\varphi    \ e^{-i p^t t +i p^\varphi \varphi} \  
         \langle  \OO_H \OO_L(t, \varphi)  \OO_L(0)  \OO_H \rangle  \, .
\ee 
In the following we will mostly set $R=1$, 
but it can be easily recovered on  dimensional grounds.
Note that the phase shift, as defined by (\ref{defps}) is related to the eikonal phase $\delta_\ell$ 
conventionally appearing in the scattering amplitudes (see e.g. \cite{Soldate:1986mk,Giddings:2007qq,Damour:2019lcq,Bern:2020gjj}) by a factor of two,
\be
\label{deltasrel}
    \delta = 2 \delta_\ell \, .
 \ee

The Fourier transformed correlator in the large $p^t,p^\varphi$ limit receives a dominant contribution from a certain
null geodesic, as we review below.
It was argued in   \cite{Kulaxizi:2018dxo}  that the phase shift is given by (\ref{psoper})
where $p^t$ and $p^\varphi$ are now the conserved quantities which determine the trajectory of the corresponding
null geodesic, while $\Delta t $ and $\Delta \varphi$ describe the deviation of the point where the null geodesic emerges at the 
boundary of the AdS-Schwarzschild from the pure AdS result.
The explicit expression for the phase shift in the $D=d+1$-dimensional AdS-Schwarzschild spacetime is  \cite{Kulaxizi:2018dxo} 
\be
\label{deltak}
\begin{split}
\delta(\sqrt{-p^2},\,L)&{=}\sum_{k=0}^{\infty}\delta_k(\sqrt{-p^2},\, L){=}\\
&{=}\sum_{k=1}^\infty {\mu^k\over k!}\,{2 \Gamma\left[{d k+1\over 2}\right]\over \Gamma\left[{k (d-2)+1\over 2}\right]}  \,{\pi^{k(d-2)+2\over 2}\over \Gamma[{k(d-2)+2\over 2}]}\,\, \,\sqrt{-p^2}\,\,\Pi_{k(d-2)+1,k(d-2)+1}(L) \,.
  \end{split}
 \ee
where
 \be
 \label{propagator}
\Pi_{\Delta - 1;d-1}(x) = { \pi^{1 - {d \over 2}} \Gamma(\Delta -1) \over 2 \Gamma(\Delta - {d -2\over 2})} \ e^{-(\Delta -1) x} \ _{2} F_1({d \over 2} - 1, \Delta - 1, \Delta - {d \over 2} + 1, e^{- 2 x})  \,. 
\ee
and
\be
\label{defL}
   e^{2 L} = {p^+\over p^-} = {p^t+p^\varphi\over p^t-p^\varphi} \, .
 \ee
 In (\ref{deltak}) and in the rest of the paper $\mu$ is proportional to the mass of the AdS-Schwarzschild black hole $M$
 and to the ratio $\Delta_H/C_T$,
 \be
 \label{defmu}
     \mu = \left[ {d-1\over 16 \pi} \Omega_{d-1} \right]^{-1} G_N M  = {4 \Gamma(d+2)\over (d-1)^2 \Gamma({d\over2}) } \ {\Delta_H \over C_T}
 \ee
 where $G_N$ is the $d+1$-dimensional Newton's constant and $\Omega_{d-1} $ is the volume of the $d-1$-dimensional sphere.

\subsection{The phase shift formula from the WKB approximation}
In this subsection we use the WKB approximation for the two-point 
function in the (thermal) CFT state dual to the AdS-Schwarzschild background  
to show that the phase shift is given by (\ref{psoper}).
(See e.g. \cite{Festuccia:2008zx} and also \cite{Balasubramanian:2019stt,Craps:2020ahu}
for examples of a WKB approximation in the computation
of a two-point funciton).

In a $d+1$-dimensional   AdS-Schwarzschild spacetime,
 \be\label{asyads}
    ds^2 = -f(r) dt^2 + {dr^2 \over f(r) } + r^2 d\Omega_{d-1} \, ,
 \ee
 with $f=1 - r^2 + \mu/r^{d-2}$, the time delay and the anglular deflection of null geodesics are given by
 \be
 \label{timeangle}
     \Delta t = 2 p^t  \int_{r_0}^\infty {dr \over f(r)  \ \sqrt{ 1 -  { f(r) \alpha^2\over r^2}  }}, \quad  \Delta \varphi = 2 p^\varphi  \alpha \int_{r_0}^\infty {dr \over r^2  \ \sqrt{ 1 -  { f(r) \alpha^2\over r^2}  }} \, ,
 \ee
 where $p^t$ ($p^\varphi$)  is the energy  (angular momentum), $\alpha =p^\varphi/p^t$
 and $r_0$ is the largest solution of $f(r ) \alpha^2 =r^2$.
  One can now use (\ref{psoper}) to arrive at the following formula for the phase shift
 \be\label{psaads}
      \delta = 2 | p^t |   \int_{r_0}^\infty {dr \over f(r)}  \ \sqrt{ 1 -  { f(r) \alpha^2\over r^2}  } \, .
 \ee

 Consider now a holographic two-point function in the thermal state   (\ref{asyads}).
 The action for a massive scalar is given by
 \be\label{acsf}
   S \simeq  \int d^5 x \sqrt{-\det g_{\alpha\beta} }\;  \left[ g^{\mu\nu} \p_\mu \phi \p_\nu \phi - m^2 \phi^2 \right] \, .
\ee
As usual, the mass is related to the conformal dimension of the dual scalar operator 
by the AdS/CFT correspondence \cite{Maldacena:1997re,Witten:1998qj,Gubser:1998bc}
 via $m^2 R^2 = \DL (\DL-d)$.
 The equation of motion is
 \be\label{eom}
    r^{1-d}  \p_r(r^{d-1} f \p_r \phi) -f^{-1} \p_t^2 \phi + r^{-2} \p_\varphi^2 \phi +m^2 \phi=0 \, .
 \ee
 Performing the Fourier transform to $\tilde\phi(p^t,p^\varphi)$,
 substituting 
 \be\label{sub}
    \tilde\phi(p^t,p^\varphi)  = e^{i p^t \psi}
\ee
  and retaining the leading terms in the 
 large energy limit, $p^t, p^\varphi \gg 1, m$, yields
  \be\label{eomeik}
      (\p_r \psi)^2 = {1\over f^2(r)}  \left(   1- {\alpha^2 f(r)\over r^2} \right) \, ,
  \ee
 which can be integrated and gives precisely (\ref{psaads}) for the phase shift.

 \subsection{Null geodesics and the phase shift}
It is not hard to prove  the following identity for the null geodesics, by performing direct differentiation
of both the lower limit of integration and the integrands,
\be
\label{usefulid}
     \alpha {\p \over \p \alpha } \int_{r_0}^\infty {\alpha dr \over r^2\sqrt{1 -  { f(r) \alpha^2\over r^2}} } - {\p\over \p \alpha }  \int_{r_0}^\infty { dr \over f(r) \sqrt{1 -  { f(r) \alpha^2\over r^2}} } =0 \, .
\ee
 With this identity, one can further show that
 \be
 \label{deltas}
     \Delta t = {\p \delta \over \p p^t} ,\qquad  \Delta \varphi = -{\p \delta \over \p p^\varphi} \, .
 \ee
Equations (\ref{deltas}) generalize the result of Appendix E in \cite{Karlsson:2020ghx}.
 Note that (\ref{deltas}) are exactly the relations between the eikonal phase and the time delay and angular deflection, familiar from
 the Regge scattering in flat spacetime 
 [they follow from the stationary phase approximation for the scattering amplitude;
 one should also bear in mind a factor of two in (\ref{deltasrel}) ].
 
 The identity (\ref{usefulid}) can  be used to illustrate how eq. (\ref{psoper}) emerges from the Fourier transform of the correlator
 in the large $\Delta_L$ limit.
 In this limit, the two-point function is related to the length of a geodesic which connects  two points on the boundary.
 At large momenta, the dominant contribution comes from   null geodesics, since they minimize the length
 (see e.g. \cite{Hubeny:2006yu} for a related discussion).
 There is one specific null geodesic which extremizes the phase shift -- its parameters can be determined by the stationary phase consition.
 We need to extremize
 \be
 \label{extrem}
    {\delta \over p^t}  = 
     \int_{r_0}^\infty {dr \over f(r)  \ \sqrt{ 1 -  { f(r) \alpha^2\over r^2}  }} - {p^\varphi\over p^t}  \alpha \int_{r_0}^\infty {dr \over r^2  \ \sqrt{ 1 -  { f(r) \alpha^2\over r^2}  }}
 \ee
with respect to the parameter  $\alpha$, which labels null geodesics.
In (\ref{extrem}) the ratio of the external momenta  $p^\varphi/ p^t$ is fixed
( $p^\varphi$ and $p^t$ are simply the variables of the Fourier transform).

According to (\ref{usefulid}) the extremum is achieved for the value of $\alpha$ which is precisely equal to 
the ratio  $p^\varphi/ p^t$, thereby confirming  (\ref{psoper}).
   
 \section{Flat space limit (small AdS impact parameter)}

\subsection{Taking the flat space limit}
It is interesting to take the flat space limit of (\ref{deltak}).
This is achieved by taking the AdS radius $R$ to be large compared to the Schwarzschild radius $R_s$ of the black hole
and the impact parameter $b$.
Recall that  \cite{Kulaxizi:2018dxo} 
\be
\label{defbmu}
   b = R \ \sinh L, \qquad \mu \approx \left( {R_s\over R} \right)^{D-3}   \, .
 \ee
Hence, the flat space limit corresponds to the limit of small $\mu$ and $L$ with
\be
\label{fixedratio}
    {\mu \over L^{D-3} }   \approx     \left( {R_s \over b} \right)^{D-3}  \approx  \left( {R_s p^t \over p^\varphi} \right)^{D-3} 
 \ee
fixed.
The result of this limit for the phase shift is
\be
\label{pslimit}
  \delta_M=  \sum_{k=1}^\infty  { \sqrt{\pi} \Gamma\left( { (D-1) k +1\over2}\right) \over  ( (D-3) k -1 ) k! \Gamma\left( { (D-3) k\over2}+1\right)}  { (R_s p^t)^{k(D-3)}   \over (p^\varphi)^{k(D-3)-1} } \, ,
 \ee
 where the subscript ''M" stands for Minkowski spacetime.
It is instructive to compute the deflection angle,
\be
\label{defdef}  \varphi_M =  -{d \delta_M \over d p^\varphi }  \, .
\ee
Differentiating eq. (\ref{pslimit}) yields
\be
\label{defa}
   \varphi_M =  \sum_{k=1}^\infty  { \sqrt{\pi} \Gamma\left( { (D-1) k +1\over2}\right) \over  k! \Gamma\left( { (D-3) k\over2}+1\right)}  \left( {R_s \over b} \right)^{k(D-3)}        
\ee
where we substituted $b=p^\varphi/p^t$.

\subsection{Four-dimensional spacetime}  

Consider the four-dimensional spacetime, $D=4$.
As an extra check, we can make use of eq. (11.32) in \cite{Bern:2019crd} (see also appendix D of \cite{KoemansCollado:2019ggb})
where the scattering angle is quoted up to the next-to-next to leading order.
To compare with our results, we need to take the probe limit, where the mass of one particle is the largest scale in the problem.
This produces
\be
\label{deffourd}
  \varphi_M = 2 \left({R_s\over b}\right)+ {15 \pi \over 16} \left({R_s\over b}\right)^2 + {16 \over 3} \left({R_s\over b}\right)^3 
     +  \ldots  \, .    
\ee
This is in complete agreement with (\ref{defa}).

It is interesting that the sum in (\ref{pslimit}) can be computed exactly.
One way to do it is to   substitute $D=4$ in (\ref{defa}) and then integrate the result. 
It will be convenient to define 
\be
\label{xdef}
     x = {3 \sqrt{3} \over2}  \ {R_s p^t\over  p^\varphi} \, ,
 \ee
which yields
\be
\label{pssum}
     \delta_M=  {3 \sqrt{3} R_s p^t \over 2 x} \left( i \pi-\pi  \ _3 F_2 [-\frac{1}{2}, \frac{1}{6},\frac{5}{6}; \frac{1}{2}, 1; x^2]  -{x \over 4} G^{2,3}_{4,4} [ -\frac{1}{3},0,\frac{1}{3},1; 0,0,-\frac{1}{2},-\frac{1}{2}; -x^2 ]       \right)+c \, ,
\ee
where $G^{2,3}_{4,4}$ is the Meijer G-function (see Appendix A) and $c$ is a real constant.
 We will be interested in the imaginary part of the phase shift, which develops for $x >1$ (this corresponds to the impact parameter of a null
 geodesic which approaches the light orbit).
  \begin{figure}
  \begin{center}
	\includegraphics[width=5in]{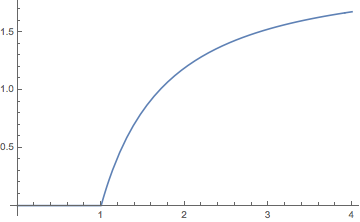}
	\caption{\label{deltaiplot} The plot of  $Im \delta_M/(R_s p^t \pi)$ }
	\end{center}
\end{figure}
 It has the form (see Appendix A)
 \be
\label{pssumim}
     Im \ \delta_M  =  R_s p^t f(x) \, ,
 \ee    
 where $f(x)$ is plotted in Fig. 1.
 The  imaginary part is vanishing (the scattering is elastic) for $x<1$ which corresponds to the impact
 parameter larger than the radius of the circular light orbit, $R_* = 3 \sqrt{3} R_s/2$.
 On the other hand, for $x>1$, $Im \ \delta_M  $ is very large (since $R_s p^t \gg 1$ in the eikonal limit we consider) -
 the scattering for these partial waves is completely non-elastic (they are totally absorbed). 
 The inelastic scattering cross-section is
 \be
 \label{xsecin}
   \sigma_{in,M} = {\pi \over (p^t)^2}  \sum_{\ell=0}^{p^t R_*}  (2 \ell+1) (1 -|  e^{i \delta}   |^2 )  = \pi R_*^2 \, .
 \ee
 This is the geometric absorption cross-section of the Schwarzschild metric.

\section{Leading twist limit (large AdS impact parameter)}

 Each term in the phase shift result (\ref{deltak}) has the following behavior in the large impact parameter regime, $L\gg 1$,
 \be\label{phaselip}
 \delta^{(k)} (p^2, L) \sim \sqrt{-p^2} e^{-(k (d-2) +1) L } \sim p^- e^{-k (d-2) L } \, ,
 \ee
 which summarizes the contributions of all leading twist k-stress tensors.
 Hence, one can take a double scaling limit $\mu\rightarrow\infty$, $\mu e^{-(d-2)  L}$ fixed, where only such 
 operators survive.

This limit  was recently  considered in \cite{Parnachev:2020fna}, where an effective three-dimensional metric 
with the boundary coordinates $x^+ = t+\varphi$, $x^- = \mu^{2\over d-2}  (t-\varphi)$ was introduced.
 It is clear that one should be able to recover (\ref{psdfour}) directly from that effective metric,
\be\label{defpsltwist}
    \delta = {1\over2} p^+ (\Delta t-\Delta \varphi  ) + {1\over2} p^- \dxp =  {p^-\over 2} \left(   {p^+ \over \mu^{2\over d-2}  p^-}  \dxm +\dxp   \right) \, .
\ee
where $\dxp \equiv \Delta t+\Delta \varphi$ and $\dxm \equiv \mu^{2\over d-2}  (\Delta t- \Delta \varphi)$
are the coordinate displacements of the null geodesic.

\subsection{Effective metric and null geodesics}
\label{sec:metric}
 We start with the AdS-Schwarschild spacetime (\ref{asyads}) and consider
the limit $\mu\rightarrow\infty$, $x^-$ fixed.
This corresponds to taking the lightcone limit and the large $\mu$ limit simultaneously.
A null geodesic  propagates in the $x^+, x^-, r$ part of the spacetime.
In the double-scaling limit the metric  (\ref{asyads}) becomes \cite{Parnachev:2020fna}
\be
\label{metlimit}
   ds^2 = -\frac{1}{4} \left(1- \frac{1}{y^2} \right) (dx^+)^2 -y^2 dx^+ \, d x^- +\frac{dy^2}{y^2} \, ,
\ee
where $y=r \mu^{-{1\over d-2}}$.
There are two conserved quantities,
 \be
 \label{conserved}
K_+=  -\frac{1}{4} \left(1- \frac{1}{y^{d-2}} \right) \dot x^+ -\frac{y^2}{2} \dot x^-, \qquad
 K_-=-K=-\frac{y^2}{2} \dot x^+ \, .
 \ee
 Note that $K>0$ and $K_+<0$ to ensure $\Delta x^\pm >0$. 
 The geodesic equation becomes
 \be\label{geo}
   \dot y^2  + 4 K K_+ +  (y^{-2}- y^{-d} ) K^2 =0 \, .
 \ee
 As usual, the problem is equivalent to a problem of one-dimensional motion in some effective potential.
We can write
\be\label{dxp}
  \dxp = 4 \int_{y_0}^\infty {  dy \over  ( y^{4-d} - y^{-2}  +\tilde \alpha  y^4 )^{1\over2}   } 
\ee
while 
\be\label{dxm}
  \dxm =  2 \int_{y_0}^\infty  dy {  y^{-d} - y^{-2} +{\tilde \alpha\over2} \over  ( y^{4-d} - y^{-2}  +\tilde \alpha y^4 )^{1\over2}   }  \, ,
\ee
where 
\be\label{alpha}
    \tilde \alpha = -{4 K_+ \over K} = {4 K_+\over K_-}= {4 \mu^{2\over d-2} p_- \over  p_+} = {4 \mu^{2\over d-2}  p^- \over p^+}  = 4 \mu^{2\over d-2}  e^{-2 L}
\ee
is kept finite in the double scaling limit.
The phase shift takes the form 
\be
\label{psleadingtw}
    \delta = {4 p^-\over \tilde \alpha} \int_{y_0}^\infty dy y^{-4} (\tilde \alpha y^4 -y^2 +y^{4-d})^{1\over2} \, .
\ee

\subsection{Leading twist in  $d=4$}
To illustrate the discussion above, consider the case of $d=4$.
One can compute the sum (\ref{deltak}) in the double scaling limit directly; the result for the leading twist phase shift is
 \be\label{psdfour}
   \delta = p^- \pi \, _2F_1 ({1\over4},{3\over 4}, 2, 16 \mu e^{-2 L} ) \, .
 \ee
One can also use the effective metric to write down the 
 explicit expressions for $\dxp$, $\dxm$:
\be\label{dxpa}
  \dxp =  {2 \pi \over \sqrt{\tilde\alpha u_0}}  \ _2 F_1({1\over2},{1\over2},1,\tilde u_1) \, ,
 \ee
\be\label{dxma}
  \dxm = {\tilde\alpha\over 4} \dxp  +{\pi \over 8 \sqrt{\alpha} }  \left(   {3\over2}  \, _2 F_1({5\over2},{1\over2},3,\tilde u_1) -\,  _2 F_1({3\over2},{1\over2},2,\tilde u_1)  \right) \, ,
 \ee
 where $u_{0,1} = (1 \pm \sqrt{1- 4 \tilde\alpha})/(2 \tilde\alpha)$ and $\tilde u_1 = u_1/u_0$.
 Now we can compute the leading twist phase shift
 \be\label{ltps}
    \delta = {2 p^-\over\tilde\alpha \sqrt{u_0}} \int_{u_0}^\infty {du \over u^{5\over2}  }   \,(\tilde \alpha u^2 - u +1)^{1\over2}  = {\pi p^-\over \sqrt{\tilde\alpha u_0}}  \, _2 F_1 (- {1\over2},{1\over2},2,\tilde u_1) \, .
 \ee 
Note that (\ref{ltps}) exactly agrees with (\ref{psdfour}), as expected (see Appendix B).

\subsection{Lyapunov exponent}
As explained in  \cite{Cardoso:2008bp} 
(and recently investigated in the context similar to that of the present paper in \cite{Bianchi:2020des,Berenstein:2020vlp})
 an interesting quantity is the Lyapunov exponent $\lambda$.
It is related to the critical behavior of $\Delta t$,
\be
\label{lyap}
    \Delta t \approx -{1\over \lambda} \log (\tilde\alpha-\tilde\alpha_c), \qquad \Delta \varphi \approx -{\omega_c\over \lambda} 
        \log (\tilde\alpha-\tilde\alpha_c) \, ,
\ee
which leads to the Lyapunov scaling as $\tilde\alpha$ approaches  $\tilde\alpha_c$,
\be
\label{lyapsca}
     {\delta \Delta \varphi \over \delta \tilde\alpha}  \sim e^{\lambda \Delta t }\, .
\ee
In the double scaling limit we consider in this Section,
$\Delta t \approx \Delta \varphi \approx \dxp/2$, and from (\ref{dxpa}) we 
infer $\tilde\alpha_c =1/4$ and $\lambda=\sqrt{2}$ for $d=4$.
More generally, one can use the explicit formula \cite{Cardoso:2008bp} to compute
\be
\label{lyapfla}
   \lambda = \sqrt{  V_{eff}''\over 2 \dot t^2} = \sqrt{d-2} \, , 
\ee
where the effective potential is read off from (\ref{geo})
and all quantities in (\ref{lyapfla}) are evaluated on the circular light orbit.

It would be interesting to investigate the relation of the classical Lyapunov exponent discussed in this section and 
the Lyapunov exponent which appears in  out-of-time ordered correlators which satisfy the bound on chaos \cite{Maldacena:2015waa}.
The latter originated from a (squared) commutator of two local operators and hence measures  quantum chaos.
The regularization used in \cite{Maldacena:2015waa}, which involves separating commutators by a half thermal circle
in  the euclidean time, leads to a correlator where operators appear on both sides of the thermofield double.
This holographically corresponds to insertions of the operators on the two sides of the ethernal AdS-Schwarzschild.
The resulting geodesic calculation  probes different geometry, compared to the one discussed in this section --
it would be interesting to see if there is a connection between the two calculations.

\section{Discussion and open questions}
In this paper we consider the eikonal phase which appears in the probe limit of
high energy gravitational scattering in AdS spacetime.
The AdS eikonal phase has been useful  in the context of  holographic CFTs, 
where it receives contributions from the stress tensor sector and is often insensitive
to the double trace contributions (see however \cite{Giusto:2020mup}).
It remains to  be seen whether it can play an important role outside of holography, since
in generic CFTs one expects a number of low lying higher spin operators, which 
would contribute to the phase shift.
It would be interesting to see how the finite gap in the spectrum of spinning operators would affect the phase shift.
This corresponds to  stringy corrections to the eikonal phase in the bulk, a subject that
received a lot of attention starting from \cite{Amati:1987wq}.
It would also be interesting to see a direct derivation of the scattering amplitude in AdS in the Regge limit.

Note that to reproduce the correct inelastic scattering cross-section of the 
Schwarzschild metric
it was sufficient to observe that the phase shift develops a large imaginary part (\ref{pssumim})
for $x>1$.
The exact behavior of the function $f(x)$ didn't matter for this conclusion, but
it would be interesting to understand it better.
Can it be obtained from some effective action for Regge scattering \cite{Lipatov:1991nf,Amati:1993tb}?
One may also wonder whether   geodesics  which probe the black hole interior (see e.g. \cite{Fidkowski:2003nf,Grinberg:2020fdj} ) 
 play a role in computing $f(x)$.

From the dual CFT point of view the flat space limit of the phase shift equals the anomalous dimension of
the corresponding heavy-light operators\footnote{It was explicitly shown to $\OO(\mu^2)$ in \cite{Karlsson:2019qfi} but we verified this to next order, and believe
 it holds generally.}
Hence, complex values of the phase shift imply complex anomalous dimensions.
Of course in the heavy-light scattering case considered here, the heavy-light operators are extremely heavy to start with.
On the other hand, the situation must be qualitatively similar for the light-light scattering.
Namely, the physics of the black hole formation at sufficiently small impact parameters should imply
 complex anomalous dimensions of the double trace operators.
 It would be interesting to see if holography can shed more light on this (see e.g. \cite{Gorbenko:2018ncu}  for a recent CFT interpretation
of complex anomalous dimensions). 
Another possibility would be a scenario  similar to what happens in a light-light scattering setup when the finite string length corrections are taken into account
and the phase shift becomes complex.
In this case new single trace operators emerge \cite{Li:2017lmh,Meltzer:2019pyl}  in the S-channel\footnote{
We thank David Meltzer for pointing this out to us.}
(heavy-light channel in our situation).

In addition to the eikonal phase, we computed the Lyapunov exponent associated with the limiting behavior 
of null geodesics as they approach the circular null orbit (the photosphere).
We obtained a universal value $\lambda = \sqrt{d-2}$ which does not depend on the addition of higher derivative
gravitational terms to the bulk action.
It is interesting to compare it with the Lyapunov exponent considered in \cite{Maldacena:2015waa}, $\lambda_C= 2\pi T$,
even though the two quantities apprarently describe different physics
($\lambda_C$ is related to the behavior of  four-point functions in the finite temperature background,
while $\lambda$ is related to the two-point function; 
in the bulk language the former is dominated by the near-horizon scattering, while
the latter reflects the behavior of geodesics near the photosphere).
The minimal value of the AdS-Schwarzschild temperature in  the units of AdS radius is $T_{min}  = (2\pi)^{-1} \sqrt{ d (d-2)  }$, which gives
$\lambda_{C,min}= \sqrt{d (d-2)} > \lambda$, so, interestingly, $\lambda$ satisfies the 
 bound on chaos \cite{Maldacena:2015waa}.

It would be interesting to see how generic the value of $\lambda$ is. 
Generalization to the asymptotically AdS black holes with rotation and/or charge should be straightforward.
Another natural question is a field theoretic interpretation of the critical behavior (\ref{lyap}).
Note that it is related to the critical behavior of the eikonal phase $\delta$, since $\Delta t$ is related to $\delta$
via (\ref{deltas}).
Presumably this critical behavior of the eikonal phase is related to the asymptotic behavior of the 
leading twist multi stress tensor OPE coefficients.
It would be interesting to make it precise.

Finally, one may  wonder whether the effective metric (\ref{metlimit}), which encodes the contributions
of leading twist multi stress operators, has  important physical significance.
It is interesting to note that this metric is not maximally symmetric and is not a solution of the vacuum Einstein
equations. 
Perhaps the asymptotic symmetries of this metric can be used to infer a higher dimensional 
analog of the Virasoro algebra\footnote{See \cite{Huang:2019fog,Huang:2020ycs}
 for  related work.} . 
We leave this for future investigation.


\bigskip
\section*{Acknowledgements}
We would like to thank M. Bianchi, A. Grillo, R. Karlsson, M. Kulaxizi, D. Meltzer, 
J.F. Morales, G-S. Ng, R. Roiban, R. Russo, C.-H. Shen, G. Sterman, P. Tadic for useful discussions, correspondence
and comments on the draft.
A.P. thanks the Aspen Center for Physics, where this project  originated, for hospitality.
This work was supported in part by the NSF grant PHY-1607611 (Aspen Center for Physics) and by the Laureate Award IRCLA/2017/82 from the Irish Research Council.

\bigskip

\appendix


\section{Analytic continuation of phase shift}
The  phase shift in $D=4$ is given by (\ref{pssum}) To continue it  to $x>1$, we use the integral expression for the functions,
\be
{}_3F_2\left[\begin{matrix}-\frac{1}{2},\frac{1}{6},\frac{5}{6}\\ \frac{1}{2},1\end{matrix};x^2\right]=-\frac{1}{4\pi}\int  ds\ \G(s)\frac{\G(-1/2-s)\G(1/6-s)\G(5/6-s)}{\G(1/2-s)\G(1-s)}(-x^2)^{-s}\,.
\ee
For $x>1$, we take the poles $s=1/6+n\,, 5/6+n$. We adjust the contour so that $s=1/2$ pole is also included. We deform the contour to exclude the pole and a clean distinction of contour. Finally,
\begin{align}
\begin{split}
{}_3F_2\left[\begin{matrix}-\frac{1}{2},\frac{1}{6},\frac{5}{6}\\ \frac{1}{2},1\end{matrix};x^2\right]_{ac}=&\frac{\G(-\frac{2}{3})\G(\frac{1}{6})}{4\sqrt[3]{2}\pi^{3/2}(-x^2)^{1/6}}\ {}_3F_2\left[\begin{matrix}\frac{1}{6},\frac{1}{6},\frac{2}{3}\\ \frac{1}{3},\frac{5}{3}\end{matrix};\frac{1}{x^2}\right]+\frac{\G(-\frac{4}{3})\G(\frac{5}{6})}{8\sqrt[3]{4}\pi^{3/2}(-x^2)^{5/6}}\ {}_3F_2\left[\begin{matrix}\frac{5}{6},\frac{5}{6},\frac{4}{3}\\ \frac{5}{3},\frac{7}{3}\end{matrix};\frac{1}{x^2}\right]\\
&-\frac{2ix}{3\sqrt{3}}\,.
\end{split}
\end{align}
Similarly the analytical continuation of the Meijer-G function is given by,
\begin{align}
\begin{split}
&G^{2,3}_{4,4}\left[-x^2\bigg|\begin{matrix}-\frac{1}{3},0,\frac{1}{3};1\\ 0,0;-\frac{1}{2},-\frac{1}{2}\end{matrix}\right]_{ac}=\pi\int ds\frac{\csc\pi s}{s}\frac{\G(2/3-s)\G(4/3-s)}{\G(3/2-s)^2}(-x^2)^{-s}\,,\\
&=\frac{\sqrt{3}\pi \G(-\frac{2}{3})}{2\G(\frac{1}{6})^2(-x^2)^{4/3}}{}_3F_2\left[\begin{matrix}\frac{5}{6},\frac{5}{6},\frac{4}{3}\\ \frac{5}{3},\frac{7}{3}\end{matrix};\frac{1}{x^2}\right]-\frac{\sqrt{3}\pi \G(\frac{2}{3})}{\G(\frac{5}{6})^2(-x^2)^{2/3}}{}_3F_2\left[\begin{matrix}\frac{1}{6},\frac{1}{6},\frac{2}{3}\\ \frac{1}{3},\frac{5}{3}\end{matrix};\frac{1}{x^2}\right]-\frac{2\sqrt{3}}{x^2}{}_4F_3\left[\begin{matrix}\frac{1}{2},\frac{1}{2},1,1\\ \frac{2}{3},\frac{4}{3},2\end{matrix};\frac{1}{x^2}\right]\,.
\end{split}
\end{align}
The relevant pole contributions being $s=1+n$, $s=2/3+n$ and $s=4/3+n$ with $n\in \mathbb{Z}_{\geq0}$. The last term in the above does not have any imaginary contribution and hence we will neglect this term subsequently. 

Hence, for $x>1$,
\begin{align}\label{xg1}
\begin{split}
{\d_M(x)\over \pi R_s p^t}&=\bigg(1+\frac{2ix}{3\sqrt{3}}-\frac{\G(-\frac{2}{3})\G(\frac{1}{6})}{4\sqrt[3]{2}\pi^{3/2}(-x^2)^{1/6}}\ {}_3F_2\left[\begin{matrix}\frac{1}{6},\frac{1}{6},\frac{2}{3}\\ \frac{1}{3},\frac{5}{3}\end{matrix};\frac{1}{x^2}\right]-\frac{\G(-\frac{4}{3})\G(\frac{5}{6})}{8\sqrt[3]{4}\pi^{3/2}(-x^2)^{5/6}}\ {}_3F_2\left[\begin{matrix}\frac{5}{6},\frac{5}{6},\frac{4}{3}\\ \frac{5}{3},\frac{7}{3}\end{matrix};\frac{1}{x^2}\right]\\
&-\frac{x}{4\pi}\bigg(\frac{\sqrt{3}\pi \G(-\frac{2}{3})}{2\G(\frac{1}{6})^2(-x^2)^{4/3}}{}_3F_2\left[\begin{matrix}\frac{5}{6},\frac{5}{6},\frac{4}{3}\\ \frac{5}{3},\frac{7}{3}\end{matrix};\frac{1}{x^2}\right]-\frac{\sqrt{3}\pi \G(\frac{2}{3})}{\G(\frac{5}{6})^2(-x^2)^{2/3}}{}_3F_2\left[\begin{matrix}\frac{1}{6},\frac{1}{6},\frac{2}{3}\\ \frac{1}{3},\frac{5}{3}\end{matrix};\frac{1}{x^2}\right]\\
&-\frac{2\sqrt{3}}{x^2}{}_4F_3\left[\begin{matrix}\frac{1}{2},\frac{1}{2},1,1\\ \frac{2}{3},\frac{4}{3},2\end{matrix};\frac{1}{x^2}\right]\bigg)\bigg)\,.
\end{split}
\end{align}
Now consider the imaginary part of the phase shift. 
For $0<x\leq1$, 
\be
\mathfrak{Im}\ \d_M(x\leq1)=0\,.
\ee
For $x>1$, 
\begin{align}
\begin{split}
\mathfrak{Im}\ {\d_M(x>1)\over  \pi R_s p^t}&=\bigg(1+\frac{2x}{3\sqrt{3}}+\frac{\G(-\frac{2}{3})\G(\frac{1}{6})}{8\sqrt[3]{2}\pi^{3/2}x^{1/3}}\ {}_3F_2\left[\begin{matrix}\frac{1}{6},\frac{1}{6},\frac{2}{3}\\ \frac{1}{3},\frac{5}{3}\end{matrix};\frac{1}{x^2}\right]+\frac{\G(-\frac{4}{3})\G(\frac{5}{6})}{16\sqrt[3]{4}\pi^{3/2}x^{5/3}}\ {}_3F_2\left[\begin{matrix}\frac{5}{6},\frac{5}{6},\frac{4}{3}\\ \frac{5}{3},\frac{7}{3}\end{matrix};\frac{1}{x^2}\right]\\
&-\frac{3\G(-\frac{2}{3})}{16\G(\frac{1}{6})^2 x^{5/3}}{}_3F_2\left[\begin{matrix}\frac{5}{6},\frac{5}{6},\frac{4}{3}\\ \frac{5}{3},\frac{7}{3}\end{matrix};\frac{1}{x^2}\right]-\frac{3\G(\frac{2}{3})}{8\G(\frac{5}{6})^2x^{1/3}}{}_3F_2\left[\begin{matrix}\frac{1}{6},\frac{1}{6},\frac{2}{3}\\ \frac{1}{3},\frac{5}{3}\end{matrix};\frac{1}{x^2}\right]\bigg)\,.
\end{split}
\end{align}
Since the hypergeometric functions and the $\G-$functions are real, the imaginary part comes only from the factor $(-)^{1\over3}$ etc. After some simplifications, 
\be
\mathfrak{Im}\ {\d_M(x>1)\over  \pi R_s p^t}=2+\frac{3\sqrt{3}}{2x}\left(\frac{9\G(\frac{1}{3})}{16\G(\frac{1}{6})^2}\frac{1}{x^{5/3}}{}_3F_2\left[\begin{matrix}\frac{5}{6},\frac{5}{6},\frac{4}{3}\\ \frac{5}{3},\frac{7}{3}\end{matrix};\frac{1}{x^2}\right]-\frac{3\G(\frac{2}{3})}{4\G(\frac{5}{6})^2}\frac{1}{x^{1/3}}{}_3F_2\left[\begin{matrix}\frac{1}{6},\frac{1}{6},\frac{2}{3}\\ \frac{1}{3},\frac{5}{3}\end{matrix};\frac{1}{x^2}\right]\right)\,.
\ee
One can show that 
\be
\lim_{x\rightarrow\infty}\mathfrak{Im}\  {\d_M(x>1)\over  \pi R_s p^t}=2 .
\ee

\section{Phase shift for leading twist}
Here we will show that (\ref{psdfour}) 
\begin{equation} 
{\d\over p^-}=\pi {}_2F_1\left(\frac{1}{4},\frac{3}{4},2,4 \a \right)
\end{equation}
exactly equals (\ref{ltps}).
Start from
\begin{equation}
{\d\over p^-}=\frac{2}{\a\sqrt{u_0}}\int_{u_0}^\infty \frac{du}{u^{5/2}}(\a u^2-u+1)^{1/2}=\frac{\pi}{\sqrt{\a u_0}}{}_2F_1\left(-\frac{1}{2},\frac{1}{2},2,\frac{u_1}{u_0}\right)\,.
\end{equation}
where, $u_{0,1}=(1\pm\sqrt{1-4\a})/(2\a)$, so that $u_0 u_1=1/\a$.  After some simplifications,
\be
{\d\over p^-}=\frac{\sqrt{2}\pi}{\sqrt{1+\sqrt{1-\r}}}{}_2F_1\left[\begin{matrix}-\frac{1}{2},\frac{1}{2}\\ 2\end{matrix};\frac{(1-\sqrt{1-\r})^2}{\r}\right]\,, \ \ \r=4\a\,.
\ee
We expand the ${}_2F_1[-1/2,1/2,2,x]$ in terms of the sum,
\be
{\d\over p^-}=\sqrt{2}\pi\sum_{k=0}^\infty \frac{(-\frac{1}{2})_k(\frac{1}{2})_k}{k!(2)_k}\frac{(1-\sqrt{1-\r})^{2k+1/2}}{\r^{k+1/2}}\,.
\ee
and,
\be
\frac{(1-\sqrt{1-\r})^{2k+1/2}}{\r^{k+1/2}}=\sum_{n=0}^\infty \frac{2^{-3/2-2k-2n}(1+4k)\G(\frac{1}{2}+2k+2n)}{n!\G(\frac{3}{2}+2k+n)}\r^{k+n}\,,
\ee
so that,
\be
{\d\over p^-}=\sqrt{2}\pi\sum_{k=0}^\infty\sum_{n=0}^\infty \frac{(-\frac{1}{2})_k(\frac{1}{2})_k}{k!(2)_k}\frac{2^{-3/2-2k-2n}(1+4k)\G(\frac{1}{2}+2k+2n)}{n!\G(\frac{3}{2}+2k+n)}\r^{k+n}\,.
\ee
We change variables $n=a-k$ with $0\leq k\leq a$. We perform the $k-$sum so that,
\be
{\d\over p^-}=\sqrt{2}\pi\sum_{a=0}^\infty \frac{2^{-1/2-2a}(2a-1/2)!}{\sqrt{\pi}a!(1+a)!}\r^a=\pi {}_2F_1\left(\frac{1}{4},\frac{3}{4},2,4\a\right)\,.
\ee

\newpage

\bibliographystyle{utphys}
\bibliography{psv3}

\end{document}